\documentclass[floatfix,aps,prx,twocolumn]{revtex4-2}

\usepackage[T1]{fontenc}
\usepackage{lipsum}  
\usepackage{graphicx}
\usepackage[english]{babel}
\usepackage{upgreek}
\usepackage{amsmath}
\usepackage{amssymb}
\usepackage{tabls}
\usepackage{dsfont}
\usepackage[colorlinks=true,citecolor=blue,linkcolor=red,urlcolor=blue]{hyperref}

\makeatletter
\def\bbl@set@language#1{%
	\edef\languagename{%
		\ifnum\escapechar=\expandafter`\string#1\@empty
		\else\string#1\@empty\fi}%
	\@ifundefined{babel@language@alias@\languagename}{}{%
		\edef\languagename{\@nameuse{babel@language@alias@\languagename}}%
	}%
	\select@language{\languagename}%
	\expandafter\ifx\csname date\languagename\endcsname\relax\else
	\if@filesw
	\protected@write\@auxout{}{\string\select@language{\languagename}}%
	\bbl@for\bbl@tempa\BabelContentsFiles{%
		\addtocontents{\bbl@tempa}{\xstring\select@language{\languagename}}}%
	\bbl@usehooks{write}{}%
	\fi
	\fi}
\newcommand{\DeclareLanguageAlias}[2]{%
	\global\@namedef{babel@language@alias@#1}{#2}%
}
\makeatother

\newcommand\varpm{\mathbin{\vcenter{\hbox{%
  \oalign{\hfil$\scriptstyle+$\hfil\cr
          \noalign{\kern-.3ex}
          $\scriptscriptstyle({-})$\cr}%
}}}}
\newcommand\varmp{\mathbin{\vcenter{\hbox{%
  \oalign{$\scriptstyle({+})$\cr
          \noalign{\kern-.3ex}
          \hfil$\scriptscriptstyle-$\hfil\cr}%
}}}}

\DeclareLanguageAlias{en}{english}

\begin{document}

\title{Crosstalk analysis for single-qubit and two-qubit gates in spin qubit arrays}

\author{Irina Heinz}
\email{irina.heinz@uni-konstanz.de}
\affiliation{Department of Physics, University of Konstanz, D-78457 Konstanz, Germany}
\author{Guido Burkard}
\email{guido.burkard@uni-konstanz.de}
\affiliation{Department of Physics, University of Konstanz, D-78457 Konstanz, Germany}


\begin{abstract}
	Scaling up spin qubit systems requires high-fidelity single-qubit and two-qubit gates. Gate fidelities exceeding $98\%$ were already demonstrated in silicon based single and double quantum dots, whereas for the realization of larger qubit arrays crosstalk effects on neighboring qubits must be taken into account. We analyze qubit fidelities impacted by crosstalk when performing single-qubit and two-qubit operations on neighbor qubits with a simple Heisenberg model. Furthermore we propose conditions for driving fields to robustly synchronize Rabi oscillations and avoid crosstalk effects. In our analysis we also consider crosstalk with two neighbors and show that double synchronization leads to a restricted choice for the driving field strength, exchange interaction, and thus gate time. Considering realistic experimental conditions we propose a set of parameter values to perform a nearly crosstalk-free CNOT gate and so open up the pathway to scalable quantum computing devices.
\end{abstract}


\maketitle

\section{Introduction}
Spin qubits \cite{Loss_1998} in silicon quantum dots \cite{Zwanenburg_2013} are a promising candidate to realize large scale quantum computers~\cite{Veldhorst_2014}. Due to the dilute nuclear spin environment and weak spin-orbit coupling silicon enables long coherence times and high fidelity spin manipulation~\cite{osti_1429799}. Single-qubit gates can be implemented via electric dipole spin resonance (EDSR) by modulating electrostatic gate voltages causing motion of the dot electrons. Two-qubit gates additionally make use of exchange interactions between neighboring electron spins \cite{PhysRevB.59.2070, Yoneda_2017, Watson_2018, Zajac_2017, PhysRevLett.107.146801} operating at a symmetric operation point (``sweet spot'') to suppress charge noise to first order~\cite{PhysRevLett.116.116801, PhysRevLett.116.110402, PhysRevLett.115.096801}. Dephasing effects can be reduced through a large energy splitting due to a strong magnetic field gradient~\cite{nichol2016highfidelity} realized by a micromagnet \cite{Yoneda_2015, Kawakami_2014}.

High fidelity CNOT gate implementations have already been proposed and demonstrated in a Si/SiGe heterostructure double quantum dot architecture  \cite{Russ_2018, Zajac_2017}. Nevertheless, scalable spin qubit platforms \cite{Zajac_2016,Ferdous_2018} suffer from unwanted interactions of qubits with the environment and fluctuations of interacting fields yielding crosstalk, dephasing and charge noise, which represent challenges for experimental realization. Hence, a better understanding of the underlying effects is crucial for error prevention and thus high fidelity performance of gates within qubit arrays. In this paper we focus on crosstalk effects of single-qubit and two-qubit gates on neighboring qubits induced, e.g., by capacitive coupling between gates, which can decrease the fidelity by several percent \cite{PhysRevX.9.021011}. Here, we concentrate on crosstalk and disregard pure gate errors affecting operating qubits, which have been studied extensively \cite{Yoneda_2017,PhysRevLett.116.116801,PhysRevX.9.021011}. 

This paper is organized as follows. For our description we consider a Heisenberg spin model in Section \ref{sec:model}. On this basis, we quantify crosstalk in terms of neighboring qubit fidelity in Section~\ref{sec:analysis} and address the responsible values to reduce crosstalk errors with simple synchronizations in Section~\ref{sec:sync}. For single-qubit rotations around the $y$-axis we also consider next nearest neighbor coupling leading to off-resonant Rabi oscillations, which also occur in systems driven with global striplines \cite{Koppens_2006, Li_2018}, and give a possible cancellation condition for unwanted rotations. Similarly we suggest a double synchronization and a thoughtful choice of values for EDSR driving and exchange interaction strengths to implement mostly crosstalk-free CNOT gates in spin qubit arrays in Section \ref{sec:CNOTsync}. In Section \ref{sec:simultan} we determine crosstalk of simultaneously driven single qubit operations which are applied in quantum algorithms and give conditions to maximize the overall fidelity. Finally, to evaluate the robustness of synchronizing Rabi frequencies we analyze the impact of charge noise on the fidelity in Section~\ref{section:noise}.
\begin{figure}[t]
	\centering
	\includegraphics[width=0.47\textwidth]{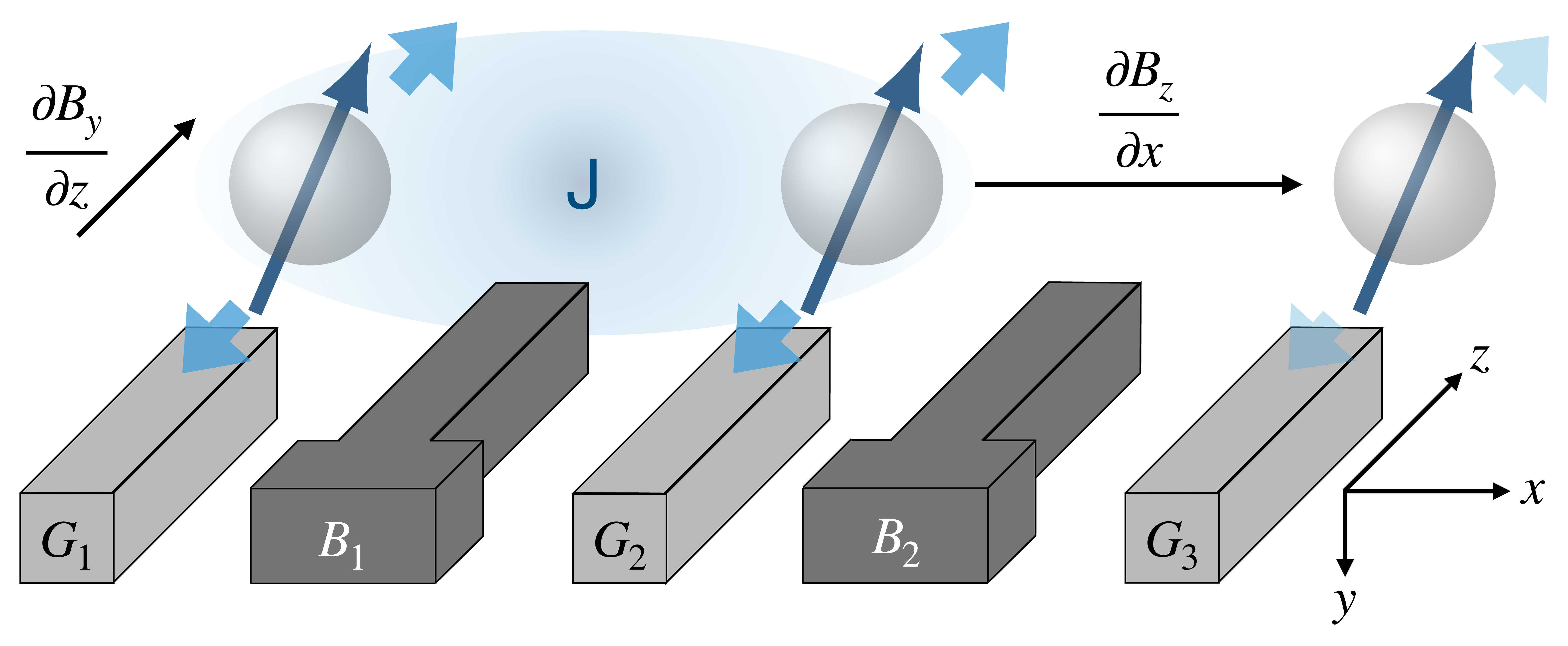}
	\caption{Schematic setup for a gate defined linear array of three quantum dots inside a large homogeneous magnetic field (not shown) and a magnetic gradient field as, e.g., induced by a micromagnet. Each dot is occupied with an electron representing a qubit with distinguishable spin resonance frequencies enabled by the gradient $\partial B_z/\partial x$. Modulations of gate voltages $G_1$ and $G_2$ shift the respective electrons in $z$-direction along the gradient $\partial B_y/\partial z$ such that an effective oscillating magnetic field is generated. This also leads to crosstalk on a neighboring qubit at gate $G_3$. Exchange interaction $J$ between two spins is controlled by electrostatic voltages at middle barrier gates $B_1$ and $B_2$ and is required for two-qubit operations.}
	\label{Qubitsystem}
\end{figure}

\section{Theoretical Model \label{sec:model}}
For our analysis we consider a gate defined linear quantum dot array  operated in the $(1,1,\ldots)$ charge regime, where the exchange interaction between two spins can be tuned by middle barrier gates, such as in Refs.~\cite{Russ_2018} and \cite{Zajac_2017} (Fig.~\ref{Qubitsystem}). 
Neglecting excited valley states and spin-orbit coupling the system can be described theoretically by the Heisenberg Hamiltonian
\begin{align}
	H = \sum_{\langle i,j\rangle} J_{ij}(t) \left( \mathbf{S}_i \cdot \mathbf{S}_j -\frac{1}{4}\right) + \sum_{i} \mathbf{S}_i \cdot \mathbf{B}_i,
\end{align}
where $J_{ij}$ is the tunable exchange interaction between nearest-neighbor spins $\mathbf{S}_i$ and $\mathbf{S}_j$, denoted by $\langle i,j\rangle$, required for two-qubit operations, and $\mathbf{B}_i = (0,B_{y,i}(t),B_{z,i})$ is the external magnetic field at the position of spin $\mathbf{S}_i$. Magnetic fields are represented in energy units throughout this paper, i.e., $\mathbf{B}_{\mathrm{physical}}=\mathbf{B}/g\mu_B$, and we furthermore set $\hbar = 1$. The total magnetic field consists of a large homogeneous field and a field gradient in $z$-direction along the $x$-axis (see Fig.~\ref{Qubitsystem}), e.g., caused by a micromagnet, $B_{z,i}=B_{z} + b_{z,i}$, allowing one to individually address single spins, and a small field gradient, such as a time dependent EDSR driving field in $y$-direction $B_{y,i}(t) = B_{y0,i} + B_{y1,i} \cos(\omega t + \theta)$. The latter in general can describe ESR or EDSR, where the effective magnetic driving strength for EDSR is proportional to the electric field depending on the device architecture, natural or artificial spin-orbit coupling mechanism, and applied gate voltage \cite{PhysRevB.74.165319, PhysRevLett.96.047202}. In the rotating frame $\tilde{H}(t) = R^{\dagger}HR+i\dot{R}^{\dagger}R$ with $R=\exp(-i \omega t \sum_{i} \mathbf{S}_i)$ we make the rotating wave approximation (RWA) which results in resonant and off-resonant Rabi terms. To evaluate crosstalk of single-qubit and two-qubit gates on the neighboring qubit we calculate the fidelity \cite{Pedersen_2007}
\begin{align}
	F = \frac{d+\left| \text{Tr}\left[ U_{\text{ideal}}^{\dagger}U_{\text{actual}} \right] \right|^2 }{d(d+1)} ,
\end{align}
where $d$ is the dimension of the Hilbert space, $U_{\text{ideal}}$ is the desired qubit operation, which in case of crosstalk would be $\mathds{1}$ for the neighboring qubits, and $U_{\text{actual}}$ the actual operation containing unwanted off-resonant Rabi oscillations with a detuned Rabi frequency
\begin{align}
\tilde{\Omega} = \sqrt{\Omega^2 + \delta\omega_{z}^2}. \label{Rabifreq}
\end{align}
Here $\Omega$ denotes the resonant Rabi frequency and $\delta\omega_z$ is the detuning between driving and resonance frequencies.

\section{Crosstalk analysis}\label{sec:analysis}
In the following, we always consider a gate performed on qubit 1 (1 and 2) for single-qubit (two-qubit) operations and their corresponding crosstalk on the nearest neighbor qubit 2 (3) in form of an unwanted  magnetic driving field on the neighboring qubit which for EDSR can be capacitively induced \cite{Cayao_2020,PhysRevApplied_12_064049} by the actual driving field $B_{y1,1}$ ($B_{y1,2}$) applied on the corresponding gate \cite{Nowack_2007}.

\subsection{Single-qubit gate: Y gate}
\begin{figure}[t]
	\centering
	\includegraphics[width=0.47\textwidth]{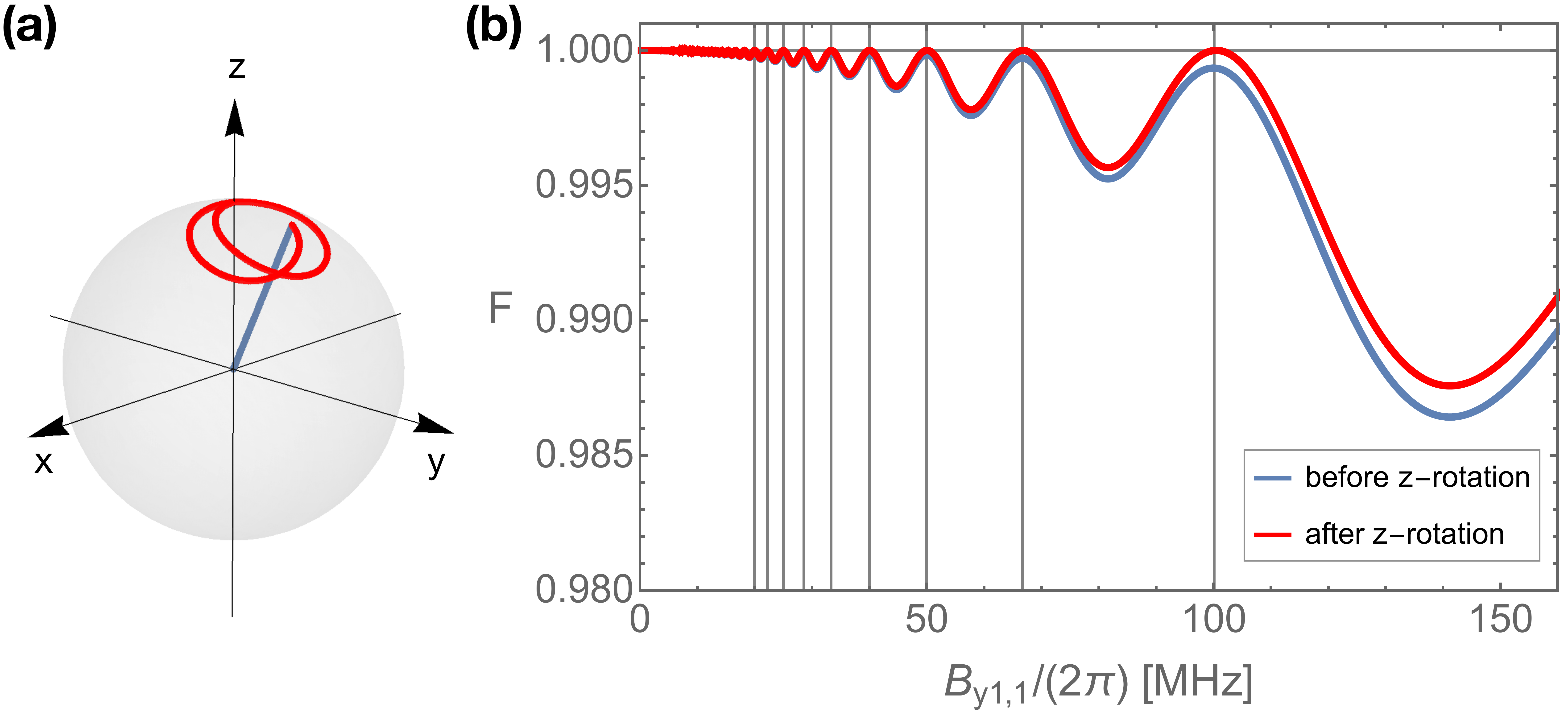}
	\caption{Fidelity of neighbor qubit 2 remaining in its state while driving a Y gate on qubit 1 at resonance frequency $B_{z,1} = (2\pi) 18.493$ GHz assuming $\alpha = 0.4$. \textbf{(a)} Off-resonant Rabi oscillations of qubit 2 in a Bloch sphere representation. \textbf{(b)} Fidelity depending on driving strength $B_{y1,1}$ (blue) shows maxima if the synchronization condition \eqref{synccond1} is fulfilled. Fidelity after subsequent $z$-rotation (red) leads to no effective crosstalk at synchronization conditions. Here $\Delta B_z = (2\pi)\, 0.2$~MHz, $B_{y1,2}=\alpha B_{y1,1}$ and  $\alpha = 0.4$.}
	\label{Ysync2}
\end{figure}

For a complete set of single qubit gates rotations around 2 axes are required. Since $z$-rotations can simply be implemented in software by incorporating the rotation angle in the phase of the microwave pulse \cite{Zajac_2017}, we focus on the performance of $y$-rotations, which are realized by driving the operating qubit 1 at its resonance frequency $\omega_{z,1}=B_{z,1}$ (a similar analysis is possible for X and other single-qubit gates). The capacitive coupling between the gate electrodes and the single electrons representing qubits can lead to a small effective magnetic field at neighboring qubit 2, which results in an off-resonant Rabi oscillation with detuned Rabi frequency Eq.~\eqref{Rabifreq}, where the resonant Rabi frequency is $\Omega_2 = B_{y1,2}/2$ and the detuning amounts to $\delta\omega_{z,2} = B_{z,2} - B_{z,1}$. For numerical examples, we assume a resonance frequency for $B_{z,1}$ of $(2\pi)18.493$~GHz as in Ref.~\cite{Zajac_2017}, and vary the nearest-neighbor qubit crosstalk field $B_{y1,2}$ as well as the $z$-field gradient $\Delta B_z = B_{z,2} - B_{z,1}$, and investigate the fidelity of the idle qubit 2. 
We find that with increasing $B_{y1,2}$ the driving strength becomes larger and the fidelity oscillates and decreases, as shown in Fig.~\ref{Ysync2}(b). On the other hand, the increasing field gradient in $z$-direction causes the resonance frequencies of the single qubits to further diverge and leads to further off-resonant Rabi oscillations such that the fidelity increases.

To verify the accuracy of the RWA used for the fidelity calculations we take into account higher-order corrections within the Floquet-Magnus expansion (FME) for periodically driven systems \cite{Bukov_2015,Blanes_2009,Moore_1990,Mostafazadeh_1997}, and 
compare RWA with FME corrected fidelities in Appendix~\ref{appFM}. Although the main contribution to the corrections comes from the non-vanishing part of $B_{y0,2}$, which is neglected in the RWA, the fidelity only slightly changes even for very large values of $B_{y0,2}$ compared to the driving field $B_{y1,2}$. Since for micromagnet-induced inhomogeneous fields the gradient field in $y$-direction is typically considerably smaller than  $B_{y1,2}$ \cite{Yoneda_2015, PhysRevB.81.085317}, we find that indeed the RWA is a good approximation for the relevant regime of operation.

\subsection{Two-qubit gate: CNOT gate}
Two-qubit gates between neighboring qubits (in our case, qubits 1 and 2) can be performed by switching on their exchange interaction as depicted in Fig.~\ref{Qubitsystem}. A CNOT gate can be implemented by adiabatically switching on $J_{12} = J$ to shift energy levels such that distinct transition frequencies allow individual addressing of the $|10\rangle\leftrightarrow |11\rangle$ transition. Combined with a driving field matching the appropriate transition frequency this results in a CNOT gate ($\Omega_{\text{CNOT}}$) with an off-resonant Rabi oscillation ($\Omega_{\text{off}}$) that can be cancelled out by a simple synchronization of Rabi frequencies \cite{Russ_2018}
\begin{align}
	\Omega_{\text{CNOT}} = \frac{2m+1}{2n} \tilde{\Omega}_{\text{off}}\hspace*{0.5cm} m, n\in\mathbb{Z}, \label{syncCNOT}
\end{align}
which we will refer to as the CNOT synchronization in the remainder of this paper. Assuming that the middle barrier gate determining $J$ has no effective capacitive coupling due to application of improved virtual gates \cite{Mills_2019, Hensgens_2017,Hsiao_2020} and therefore does not affect the neighboring qubits, the remaining effect contributing to crosstalk is a resulting driving field at the neighboring qubit 3. This turns out to be the same effect as for single-qubit gates with different driving frequency $\omega_{\text{CNOT}} = (B_{z,1} + B_{z,2} + J - \sqrt{(B_{z,2}-B_{z,1})^2+J^2})/2$, and thus different off-resonant Rabi-oscillations due to the detuning $\delta \omega_z = B_{z,3} - \omega_{\text{CNOT}}$. Calculating the fidelity depending on the resulting driving field $B_{y1,3}$ at neighboring qubit~3, the magnetic field gradient $\Delta B_z$ indeed shows a similar behavior to the case of single-qubit gate crosstalk (Fig.~\ref{Ysync2}). Since we operate in the $J\ll \Delta B_z$ regime, varying the exchange interaction $J$ only leads to small-amplitude oscillations of the fidelity.

\section{Schemes to avoid crosstalk}\label{sec:sync}
Now that crosstalk of single-qubit and two-qubit gates can be quantified, we suggest schemes minimizing these effects to reduce errors when scaling up qubit architectures beyond existing dynamical decoupling protocols~\cite{PhysRevB.97.045431,blanvillain2012suppressing}.

Off-resonant Rabi oscillations caused by driving a neighboring qubit to perform Y or CNOT gates are combined rotations around the $y$- and $z$-axes as shown in the Bloch sphere representation in Fig.~\ref{Ysync2}(a). Since $z$-rotations can be cancelled easily via software, we can find a subsequent rotation to minimize crosstalk, and thus to maximize the fidelity. For the Y gate fidelities are compared to those after following $z$-rotations as in Fig.~\ref{Ysync2}(b), and indeed, this results in a better performance. Similarly, fidelity improving $z$-rotations can also be found for two-qubit gates such as the CNOT.

In analogy to virtual gates for dc voltages defined by capacitance matrices \cite{Hanson_2007, Volk_2019} one could think of a similar approach for an ac drive, where neighboring gates would have opposite driving amplitudes but same frequency. This would cancel out driving amplitudes on neighboring qubits, and so would avoid crosstalk effects. However, the ac virtual gate strategy requires high precision control in experiments and could lead to further noise through fluctuating magnetic driving amplitudes.

A different approach, which can also be relevant for systems driving qubits with a global stripline as in \cite{Koppens_2006} and \cite{Li_2018}, is the synchronization of Rabi frequencies, similar to the one in Eq.~(\ref{syncCNOT}) of Ref.~\cite{Russ_2018}. We require that during the driving time $\tau$ to perform spin rotations the neighboring spin shall do only full $2\pi$ rotations,
\begin{align}
	\tilde{\Omega} \tau = 2\pi k, \hspace*{0.5cm} k\in\mathbb{Z}, \label{generalCROSSTALKsync}
\end{align}
with the detuned Rabi frequency $\tilde{\Omega}$. Further defining 
\begin{align}
	\alpha = \frac{B_{\text{NN}}}{B_{\text{drive}}},
\end{align}
as the ratio between the induced driving amplitude $B_{\text{NN}}$ at the nearest neighbor qubit 2 (3) and the actual driving strength $B_{\text{drive}}$ on qubit 1 (1 and 2), a condition for $B_{\text{drive}}$ can be expressed in terms of $\alpha$.

\subsection{Synchronization for the Y gate}\label{sec:Ysync}
In the case of the Y gate where $B_{\text{drive}}=B_{y1,1}$ and $B_{\text{NN}}=B_{y1,2}$ we obtain $\tilde{\Omega} = \tilde{\Omega}_2 = \sqrt{ (B_{y1,2}/2)^2 + ( \Delta B_{z})^2}$ for the off-resonant Rabi frequency of qubit 2. Since the time to perform a $\pi$ rotation on qubit 1 is $\tau_{\text{Y}} = \pi(2m+1)/\Omega_1$ where $m\in\mathbb{Z}$ with resonant Rabi frequency $\Omega_1 = B_{y1,1}/2$, we obtain
\begin{align}
	B_{y1,1} = \frac{2\Delta B_z}{\sqrt{\frac{4 k^2}{(2m+1)^2}-\alpha^2}}, \label{synccond1}
\end{align}
as synchronization condition for integer $k$ and $m$. For $\alpha = 0$ the remaining oscillation is around the $z$-axis and can be neglected, for $0 < \alpha < 1$ the fidelity reaches a maximum when condition \eqref{synccond1} is fulfilled. In Fig.~\ref{Ysync2}(b) for $\alpha = 0.4$ the fidelity is plotted in dependence of $B_{y1,1}$. Repeated fidelity maxima occur when the synchronization condition \eqref{synccond1} is satisfied. The red line of Fig.~\ref{Ysync2}(b) contains a subsequent $z$-rotation, which at the maxima in fact leads to fidelities of 1, and thus to the complete absence of crosstalk effects.

\begin{figure}[ht]
	\centering
	\includegraphics[width=0.47\textwidth]{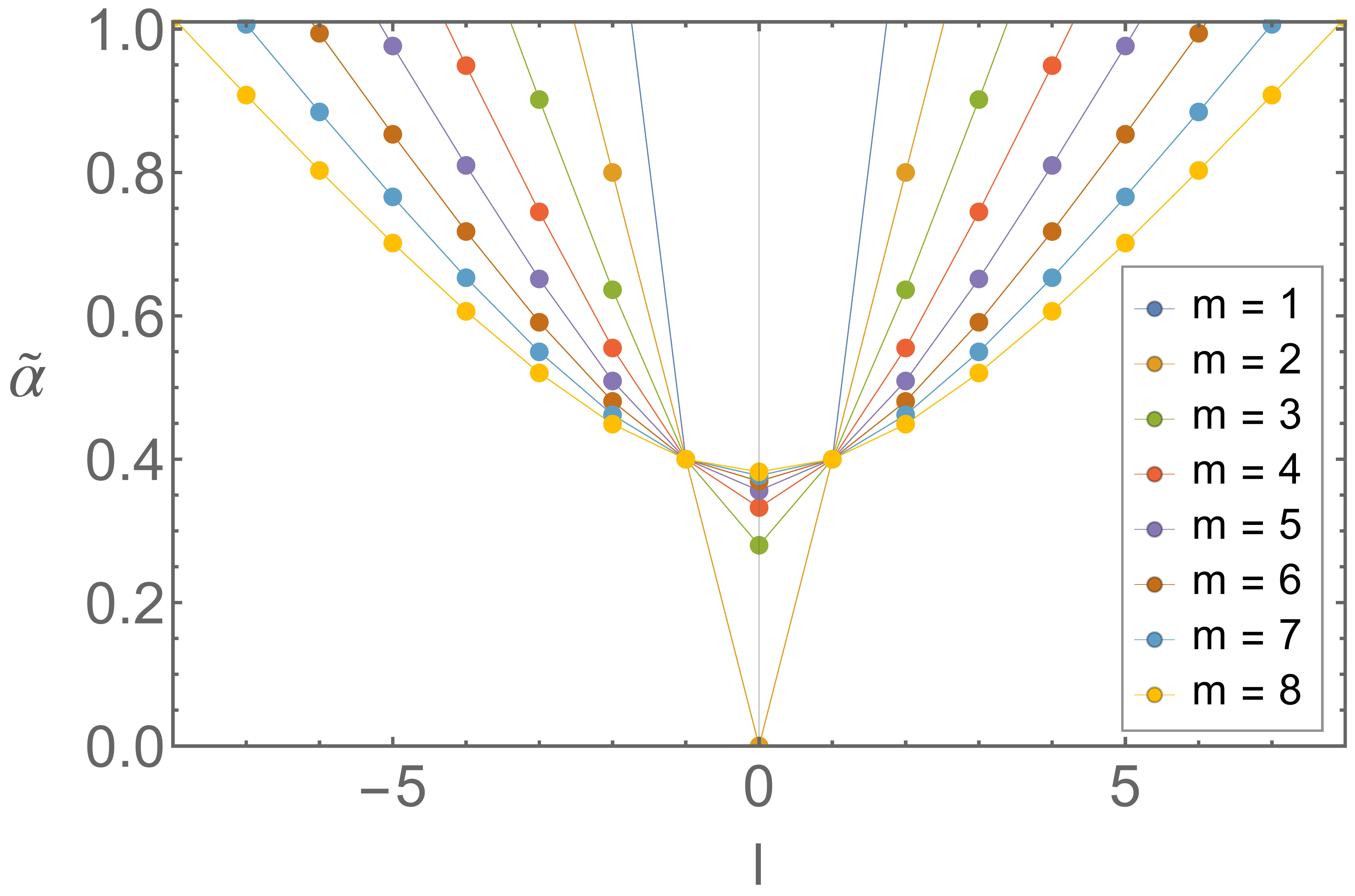}
	\caption{Solutions $\tilde{\alpha}$ for combinations of $l$ and $m$ matching the synchronization condition \eqref{synccond1} of a Y gate for two nearest neighbor qubits with $k=1$ and $\alpha=0.4$.}
	\label{YsyncNNN}
\end{figure}

So far we only considered one neighboring qubit next to the operation qubit. However, in quantum dot arrays each qubit has multiple neighbors, e.g. two neighbors in a linear array. For the Y gate we now consider two neighboring qubits 2 and 3,  with crosstalk amplitudes $B_{y1,2}=\alpha B_{y1,1}$ and $B_{y1,3}=\tilde{\alpha} B_{y1,1}$, respectively. Assuming the same gradient in z-direction for each neighbor $|B_{z,3} - B_{z,1}|=|B_{z,2} - B_{z,1}|=\Delta B_z$  condition \eqref{synccond1} must be fulfilled for each neighbor with $\alpha$ or $\tilde{\alpha}$ and integers $k$ and $l$, where $l$ replaces $k$ in condition \eqref{synccond1} for qubit 3. Since both conditions need to be simultaneously satisfied, this leads to restricted solutions for $\tilde{\alpha}$ depending on $\alpha$, $k$, $l$ and $m$. To showcase some configurations for which such a double synchronization is possible we choose $\alpha = 0.4$ and $k=1$ and represent solutions for $\tilde{\alpha}$ depending on $l$ for several values $m$ in Fig.~\ref{YsyncNNN} by dots. Besides trivial solutions we find only discrete values for $0<\tilde{\alpha}<1$. Additionally varying $k$ allows for further discrete values for $\tilde{\alpha}$. As a consequence this means that if hardware can be implemented precisely enough to determine the capacitive couplings $\alpha$ and $\tilde{\alpha}$, such that they match both synchronization conditions, crosstalk on both neighbors can be prevented completely. Alternatively, one could adapt this condition to synchronize next nearest neighbor qubits within a quantum dot array by modifying $\alpha$ and $\tilde{\alpha}$ and the gradient field respectively. Note that higher $m$ values have a direct impact on the gate time due to the $\tau_{\text{Y}} \propto 2m+1$ scaling of a $\pi$-pulse.

In general each qubit needs to be individually characterized, and differs from others due to fabrication. Therefore the above assumption for equal z-gradient fields must be adapted. Typically, $\alpha$ and $\tilde{\alpha}$ cannot be chosen but are predetermined by the system and can be obtained, e.g., by measuring frequencies of off-resonant Rabi oscillations. Although a general exact double synchronization for arbitrary $\alpha$ and $\tilde{\alpha}$ is not possible, crosstalk reduction can still be realized by choosing a configuration of $k$, $l$ and $m$, such that the synchronisation condition is nearly fulfilled.

\subsection{Synchronization for the CNOT gate}\label{sec:CNOTsync}
To perform a high fidelity CNOT gate on qubit 1 and 2 the synchronization of Rabi frequencies in Eq.~\eqref{syncCNOT} is advantageous. Similar to previous double synchronization, we would like to find solutions, which additionally satisfy the crosstalk cancellation condition \eqref{generalCROSSTALKsync}. As shown in Ref.~\cite{Russ_2018} for equal driving fields $B_{y1,1} = B_{y,1,2}$ the CNOT synchronisation yields
\begin{align}
	B_{y1,1} = \pm\frac{2J}{\sqrt{\frac{4n^2}{(2m+1)^2} \left( 1+\frac{J}{2\Delta B_{z}} \right) ^2 - \left( 1-\frac{J}{2\Delta B_{z}} \right) ^2}}, \label{CNOTBcond}
\end{align}
and thus offers $J$, $n$ and $m$ as free parameters, noting that $J~\ll~\Delta B_{z}$. In condition \eqref{generalCROSSTALKsync} we replace the off-resonant Rabi frequency $\tilde{\Omega}$ with $\tilde{\Omega}_3 = \sqrt{(B_{y1,3}/2)^2 + (B_{z,3}-\omega_{\text{CNOT}})^2}$ and the driving time with $\tau_{\text{CNOT}} = |2\pi(2m+1)/(B_{y1,1}(1+J/(2\Delta B_{z})))|$, and obtain the condition
\begin{align}
	\frac{(2m+1) \sqrt{\left( \frac{B_{y1,3}}{2}\right) ^2 + \left( B_{z,3} - \omega_{\text{CNOT}} \right) ^2}}{B_{y1,1} \left( 1+\frac{J}{2\Delta B_z} \right) } = k, \label{CNOTcond2}
\end{align}
for arbitrary $k\in \mathbb{Z}$.
This leads to solutions for $J$ depending on $k$, $n$, $m$ and $\alpha = B_{y1,3}/B_{y1,1}$. To assure that $J\ll \Delta B_{z}$ holds and the assumptions made for the CNOT gate are valid we only consider solutions with $|J|\le (2\pi)\, 20$ MHz. We find that real-valued solutions only exist for $m=0$ or $m=-1$ and we choose $m=0$ from now onward. 

\begin{figure}[ht]
	\centering
	\includegraphics[width=0.47\textwidth]{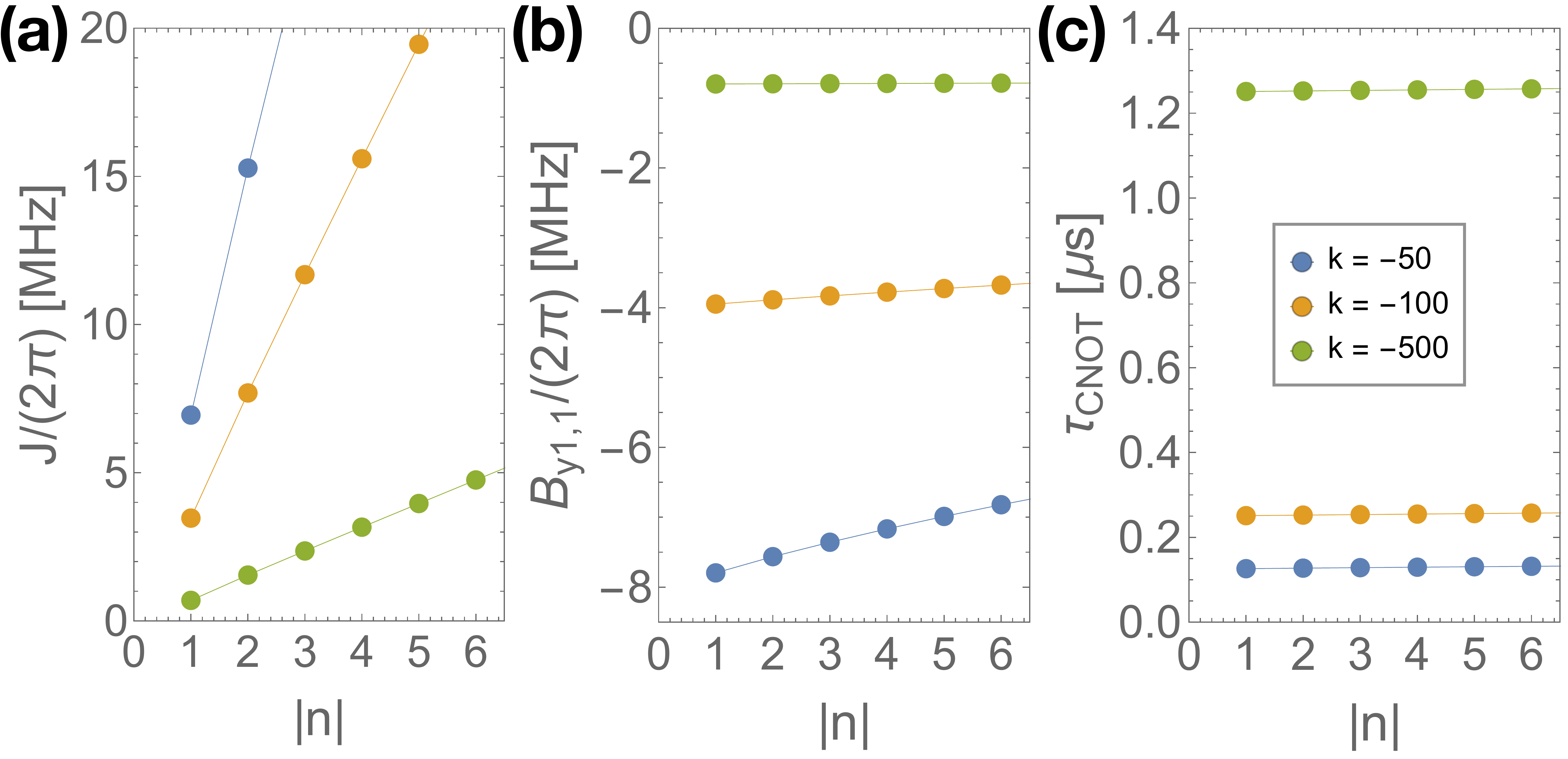}
	\caption{Symmetric solutions for exchange coupling $J$, magnetic driving amplitude $B_{y1,1}$ and driving time $\tau_{\text{CNOT}}$ depending on $n$ for $k=-50$, $-100$ and $-500$ to simultaneously fulfill the CNOT and crosstalk synchronization conditions. Here, $B_{z,1} = (2\pi)\, 18.493$~GHz, $\Delta B_{z} = (2\pi)\, 0.2$~GHz and $\alpha = 0.1$.}
	\label{CNOTsync2}
\end{figure}

In Fig.~\ref{CNOTsync2}, our results for $J$, $B_{y1,1}$ and $\tau_{\text{CNOT}}$ are shown for various values of $n$ and for $k=-50$, $-100$, and $-500$ assuming $\alpha = 0.1$. 
The general trend in the plotted region is a linearly increasing positive value for $J$ when $n$ increases. Furthermore, when choosing higher $k$ more solutions emerge in the $(2\pi)$20~MHz regime, and hence, we obtain a decreasing exchange interaction for an increasing absolute integer of $k$. On the other hand when $n$ or $k$ increase, the absolute value of $B_{y1,1}$ decreases. For larger $k$ the slope and thus the impact of $n$ is reduced. As a consequence of large driving amplitudes, due to $\tau_{\text{CNOT}} \propto |1/B_{y1,1}|$ longer gate times are required, which is why the choice of small $n$ and $k$ seem advantageous for our purpose. Nevertheless, the choice of $k$ dominates the driving time for large integers $k$, which is why $n$ is negligible for $\tau_{\text{CNOT}}$. Moreover, better validity of the approximation using small $J$ compared to the magnetic field gradient is possible at the cost of a longer gate time. Here it also turns out that a larger ratio between induced and driving field does not increase $J$ significantly. Although $\alpha$ determines the strength of crosstalk, it only slightly changes the conditioned values for synchronization due to high $k$ values.

To provide an example, for the measured resonance frequencies $B_{z,1} = (2\pi)\, 18.493$~GHz, $B_{z,2} = (2\pi)\, 18.693$ GHz, $B_{z,3} = (2\pi)\, 18.893$~GHz and $\alpha = 0.1$, we propose to set $k=50$, $m=0$, $n=1$, and thus choose an exchange coupling $J= (2\pi)\, 7.0$~MHz and driving strength $B_{y1,1} = B_{y1,2} = (2\pi)\, 7.8$~MHz, which lead to a driving time of approximately 126~ns, to fulfill both CNOT and crosstalk synchronization conditions, and hence avoid crosstalk on nearest neighbors. The precision of these values is chosen sufficiently high such as to allow for a infidelity below  $10^{-6}$.

\section{Simultaneously driven single qubit gates}\label{sec:simultan}
In quantum algorithms single qubit gates are usually applied simultaneously due to the limited coherence times of qubits. Motivated by recent crosstalk measurements in Refs.~\cite{fedele2021simultaneous} and \cite{PhysRevX.9.021011, Nature555.25766}, we further consider simultaneously driven Y gates on each of the two neighboring qubits 1 and 2, and investigate crosstalk effects. The second driving strength enters the Hamiltonian via the magnetic field in $y$-direction, which thus becomes $B_{y,i}(t) = B_{y0,i} + B_{y1,i} \cos(\omega_1 t + \theta_2) + B_{y2,i} \cos(\omega_2 t + \theta_2)$. Here we assume a symmetric crosstalk, i.e. $B_{y1,2} = \alpha B_{y1,1}$ and $B_{y2,1} = \alpha B_{y2,2}$. In the rotating frame defined by the transformation $\tilde{R}=\exp(-i (\omega_1 \mathbf{S}_1 + \omega_2 \mathbf{S}_2 ) t)$, where $\omega_1$ and $\omega_2$ match the resonance frequencies of qubit 1 and 2, the RWA is invalid for close off-resonant Rabi oscillations with large driving amplitudes $B_{y2,1}$ and $B_{y1,2}$. In this case, time dependent terms remain and lead to the $2\pi/\Delta B_z$ periodic Hamiltonian
    \begin{align}
	\tilde{H} =& \frac{1}{2} \begin{pmatrix}B_{y2,1} \sin(\Delta B_{z} t)\\B_{y1,1} + B_{y2,1} \cos(\Delta B_z t)\\0\end{pmatrix}\cdot \mathbf{S}_1\nonumber\\ 
	&+ \frac{1}{2} \begin{pmatrix}  B_{y1,2} \sin(\Delta B_{z} t)\\B_{y2,2} + B_{y1,2} \cos(\Delta B_z t)\\0 \end{pmatrix}\cdot \mathbf{S}_2.
\end{align}
An approximated description of the time evolution is given by the FME with up to second order coefficients showing a periodicity with periods $T_{\text{q}\beta}$ given by
    \begin{align}
     \left|\frac{ 16\pi\Delta B_z}{\sqrt{4 B_{y\beta,\beta} B_{y\bar\beta,\beta}^3 - 4 B_{y\beta,\beta}^2 B_{y\bar\beta,\beta}^2 - 16 B_{y\beta,\beta}^2 \Delta B_z^2 - B_{y\bar\beta,\beta}^4}}\right|, 
    \label{t} 
\end{align}
where $\beta=1,2$ denotes the qubit number and $\bar 1 =2$ and $\bar 2 =1$. During the gate time $\tau_{\text{Y}}$ both fields and their induced crosstalk fields on the neighbor are present. Choosing $B_{1,1}$ such that a $\pi$-rotation of qubit~1 is fulfilled, i.e. $\tau_{\text{Y}} = (m_1+1/2)T_{\text{q}1}$ with integer $m_1 \in \mathbb{Z}$, the single qubit fidelity of qubit~1 reaches a maximum when additionally $\tau_{\text{Y}} = 2\pi k/\Delta B_z$ with integer $k \in \mathbb{Z}$ is fulfilled. Since $\tau_{\text{Y}}$ grows linearly but fidelities increase with $k$ we choose $k=20$ which for $\Delta B_z = (2\pi)\, 0.2$~GHz leads to a gate time of 100 ns, and set $m_1 = 0$ from now on. 
\begin{figure}[t]
	\centering
	\includegraphics[width=0.47\textwidth]{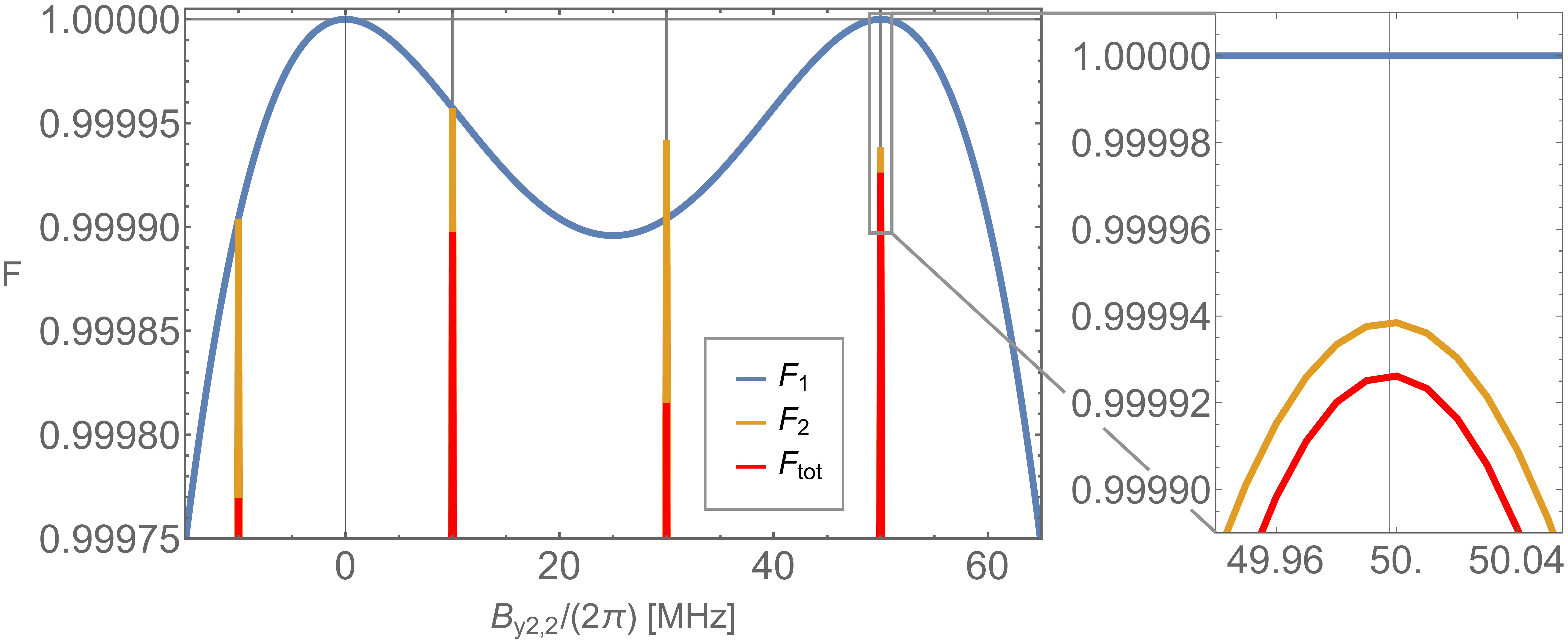}
	\caption{Single qubit fidelities $F_1$, $F_2$ of qubit 1 (blue) and 2 (yellow), and two qubit fidelity $F_{\text{tot}}$ (red) for simultaneous Y gates on qubit 1 and 2 depending on $B_{y2,2}$, while $B_{y1,1}$ fulfills $(m_1+1/2)T_{\text{q}1} = 2\pi k/\Delta B_z$. $F_1$ shows a maximum which is close to a peak of $F_2$ at $B_{y2,2}$ values fulfilling $(2+1/2)T_{\text{q}2} = 2\pi k_2/\Delta B_z$, and thus $F_{\text{tot}}$ becomes maximal.}
	\label{fig:simultan}
\end{figure}
Varying $B_{y2,1}$ shows a second fidelity maximum besides the trivial one for qubit~1 as in Fig.~\ref{fig:simultan} (blue), which occurs if $B_{y1,1}$ becomes maximal within both conditions for $\tau_{\text{Y}}$. Advantageously, for $\alpha = 0.4$ this maximum appears close to $B_{y2,2}$ fulfilling $\tau_{\text{Y}} = (m_2+1/2)T_{\text{q}2}$ with $m_2 = 2$, which is given at the maxima of the qubit 2 fidelity (yellow). Also the two qubit fidelity in Fig.~\ref{fig:simultan} (red) indeed becomes maximum at this point. Due to the assumption of symmetric crosstalk the same behaviour is obtained when swapping qubits 1 and 2. For arbitrary $\alpha$ a combination of $m_1$, $m_2$ and $k$ can be found to bring the maxima of both single qubit fidelities close together and thus nearly fulfill all the above conditions for the magnetic fields and driving time. For $B_z = (2\pi)\,18.493$~GHz, $\Delta B_z = (2\pi)\, 0.2$~GHz, $\alpha = 0.4$, $m_1=0$ and $k=20$ we find $1-F_1=3.54 \cdot 10^{-12}$ and $1-F_2=6.14 \cdot 10^{-5}$ for the single qubit infidelities of qubit 1 and 2 and $1-F=7.37 \cdot 10^{-5}$ for the two qubit infidelity where $B_{y1,1} = (2\pi)\, 10$~MHz and $B_{y2,2} = (2\pi)\, 50$~MHz. Subsequent single qubit $z$-rotations or continuing driving one of the qubits does not improve the fidelity. Hence, we obtain an unavoidable crosstalk effect for simultaneously driven spin qubits. However, to further suppress this effect in linear spin qubit chains, we propose a three-step operation to perform single qubit operations simultaneously. In each step only every third qubit is acted on at the same time, such that after three steps each qubit has been active once. For each step the driving strength can be chosen such that one neighbor qubit is synchronized by the condition~\eqref{generalCROSSTALKsync} and the other neighbor is nearly synchronized as discussed in Section~\ref{sec:Ysync}. This way crosstalk errors become minimal and the time for simultaneously performing single qubit rotations scales with a fixed factor of 3.

\section{Charge noise analysis}\label{section:noise}
Fluctuations of electric field amplitudes in silicon spin qubit devices represent a great challenge for coherence times and unitarity of control sequences \cite{RevModPhys.86.361}. Charge noise couples to spins via spin orbit coupling, magnetic field gradients and exchange coupling, which determine detuning, and thus Rabi frequencies in a CNOT gate. Since the impact of charge noise via EDSR is rather small compared to the exchange coupling \cite{Yoneda_2017}, usually fluctuations $\delta J$ of $J$ are considered. This dominant charge noise contribution does not appear for single-qubit rotations and only affects the CNOT synchronization in Eq.~\eqref{syncCNOT}, and thus the CNOT gate fidelity, which was already discussed in Ref.~\cite{Russ_2018}.
The crosstalk synchronization condition \eqref{generalCROSSTALKsync} for a neighboring qubit depends on its off-resonant Rabi frequency and driving time, and hence driving field and detuning. Since the neighboring qubit does not interact with other qubits, both driving and detuning are independent of $J$ fluctuations, and so the synchronization to avoid crosstalk is to a large extent unaffected by charge noise. 

\begin{figure}[t]
	\centering
	\includegraphics[width=0.47\textwidth]{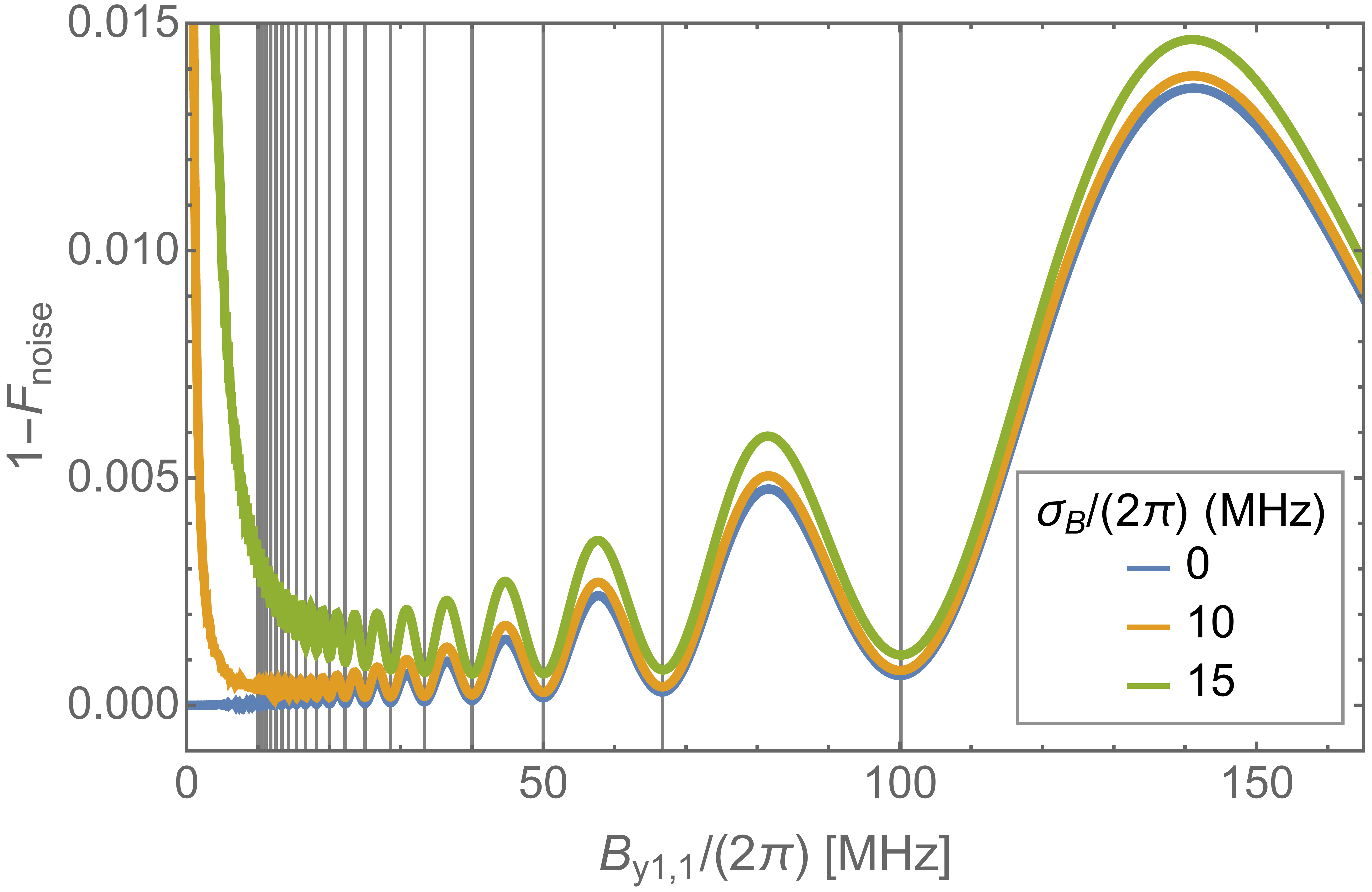}
	\caption{Total memory infidelity in the presence of crosstalk with first order noise correction in the Hamiltonian of neighboring qubit 2 for the Y gate as a function of the driving field strength $B_{y1,1}$ for three standard deviation values $\sigma_{B}$ of the noise in the effective magnetic field, where $\delta\omega_z = (2\pi)\, 0.2$ GHz and $\alpha = 0.4$. Vertical lines represent driving fields fulfilling the condition \eqref{generalCROSSTALKsync}.}
	\label{noise}
\end{figure}

Here, we consider fluctuations of driving amplitudes due to charge noise during EDSR, where the impact on crosstalk synchronization is no longer negligible. Assuming a precise driving time and non-fluctuating driving frequencies we consider the diagonal and off-diagonal time evolution terms $U_{\text{diag}} = \cos(\tilde{\Omega}_{NN}t/2) 
\mp i (\delta\omega_z/\tilde{\Omega}_{NN}) \sin(\tilde{\Omega}_{NN}t/2) $ and $U_{\text{off-diag}} = \pm (\Omega_{\text{NN}}/\tilde{\Omega}_{\text{NN}}) \sin(\tilde{\Omega}_{\text{NN}}t/2)$ of the neighboring qubit describing Rabi oscillation and calculate the first order correction in Appendix~\ref{appNoise}, which for the synchronization condition \eqref{generalCROSSTALKsync} holds
\begin{align}
    U_{\text{diag}} = 1 + i\left| \frac{\pi(2m+1)\alpha}{4 \delta \omega_z} \left( 1- \frac{(2m+1)^2\alpha^2}{4 k^2}\right) \right| \delta B,\\
    U_{\text{off-diag}} = \left| \frac{(2m+1)^3\pi \alpha^2}{16 k^2 \delta\omega_z} \sqrt{\frac{4k^2}{(2m+1)^2\alpha^2} -1}\right| \delta B,
\end{align}
where $\delta B$ denotes the fluctuation in the effective magnetic field caused by charge noise.
Accordingly, Rabi oscillations are sensitive to first order fluctuations of $B_{\text{NN}}$, which arise e.g. through charge fluctuations in the EDSR driving field. Nevertheless, applying the synchronization condition yields terms with $(2m+1)^2/k$ and $(2m+1)^3/k^2$ dependence and thus makes it possible to reduce first order noise due to short driving times and high $k$ values. The major effect of noise is obtained for the imaginary diagonal part and can be reduced by using short driving times and larger detuning $\delta \omega_z$. Furthermore, numerical evaluation of our system with $\delta\omega_z = (2\pi)\, 0.2$ GHz and $\alpha = 0.4$ shows that zeros of the first order noise contribution to the off-diagonal terms are close to the synchronization condition \eqref{generalCROSSTALKsync} for the Y gate. Since diagonal relative noise contributions are smaller than those for the off-diagonal terms, noise becomes minimal at these points. In Fig.~\ref{noise} the total infidelity of first order corrections in the Hamiltonian ($B_{y1,2} \rightarrow B_{y1,2} + \delta B$) for a zero-mean Gaussian error with standard deviations $\sigma_{B}$ of 0 MHz, $(2\pi)\,$10 MHz and $(2\pi)\,$15 MHz are shown, where one indeed finds noise to be minimal close to magnetic driving fields $B_{y1,1}$ fulfilling the condition \eqref{generalCROSSTALKsync}.
Similar behaviour is obtained for the CNOT gate implementation. Consequently, we find the crosstalk synchronization to be robust under charge noise.

Advantageous design enables qubits to be only higher order sensitive to charge noise by operating at the sweet spot \cite{PhysRevLett.116.116801, PhysRevLett.116.110402, PhysRevLett.115.096801}, which mostly reduces the above discussed effects. The main reduction in fidelity for the CNOT gate operation caused by $\delta J$ can be further decreased by applying advanced pulse shaping \cite{RevModPhys.76.1037} or dynamical decoupling sequences \cite{blanvillain2012suppressing}.

\section{Conclusions \label{sec:conclusion}}
In this paper we have analysed the effect of crosstalk of single-qubit and two-qubit gates on spin qubits and its impact on the fidelity starting from a simple Heisenberg model. We clarified the validity of the RWA in our system and are able to determine a fidelity dependence for magnetic field gradients in $y$-direction which can invalidate the RWA. Here we explicitly considered rotations around the y-axis, which together with frame-rotations around the z-axis enable arbitrary single-qubit operations. Since dc driven two-qubit gates as the CPHASE and $\sqrt{\text{SWAP}}$ only require a square pulse, their crosstalk effects can mostly be avoided by virtual gates. Here we consider a CNOT as two-qubit gate, which is operated in the $J\ll \Delta B_z$ regime and driven with an ac pulse. We showed that the underlying effect for crosstalk is the same for single-qubit and ac driven two-qubit gates, namely a residual driving field at neighboring qubits leading to off-resonant Rabi oscillations. 

Moreover, we considered techniques to minimize crosstalk besides existing dynamical decoupling schemes and suggested a virtual gate realization for driving fields. We also demonstrated how software implementations of subsequent $z$-rotations can significantly increase the fidelity. Then we proposed a synchronization of Rabi frequencies to avoid crosstalk and gave synchronization conditions for the single-qubit Y gate and the two-qubit CNOT gate. This idea is in general applicable to any device suffering from crosstalk via off-resonant Rabi oscillations, and only requires the knowledge of the off-resonant Rabi frequencies and precise control of driving strength and time. Furthermore, the synchronization protocol can also be adapted to partial $\pi$-rotations, and thus is compatible with spin echo pulsing leading to even higher fidelities in experimental setups. We also showed that double synchronization with two neighboring qubits is possible in principle, under some additional constraints, and requires precise hardware fabrication and knowledge to exactly determine the capacitive coupling between driving gates. For multiple neighbors of an operating qubit with arbitrary couplings as in qubit arrays an exact synchronization of one qubit, e.g. the most affected one, is possible and approximate synchronizations of the other neighbors can be found. For a detailed solution for a given number of neighbouring qubits further work is required and goes beyond the scope of this paper. We gave realistic exchange interactions and driving strengths to perform mostly crosstalk-free Y and CNOT gates and thus suggested a robust high fidelity gate implementation for scaled-up systems containing multiple quantum dot spin qubits. For simultaneously driven single qubit rotations we found fidelity maximizing conditions for driving time and strengths and suggested a three-step application of simultaneous gates to reduce and control crosstalk, and therefore speed up and improve quantum algorithms on spin qubit devices.

\section*{Acknowledgments}
This work has been supported by QLSI with funding from the European Union's Horizon 2020 research and innovation programme under grant agreement No 951852 and by the Deutsche Forschungsgemeinschaft (DFG, German Research Foundation) Grant No. SFB 1432 - Project-ID 425217212.

\appendix

\section{Floquet-Magnus expansion}\label{appFM}
\begin{figure}[b]
	\centering
	\includegraphics[width=0.47\textwidth]{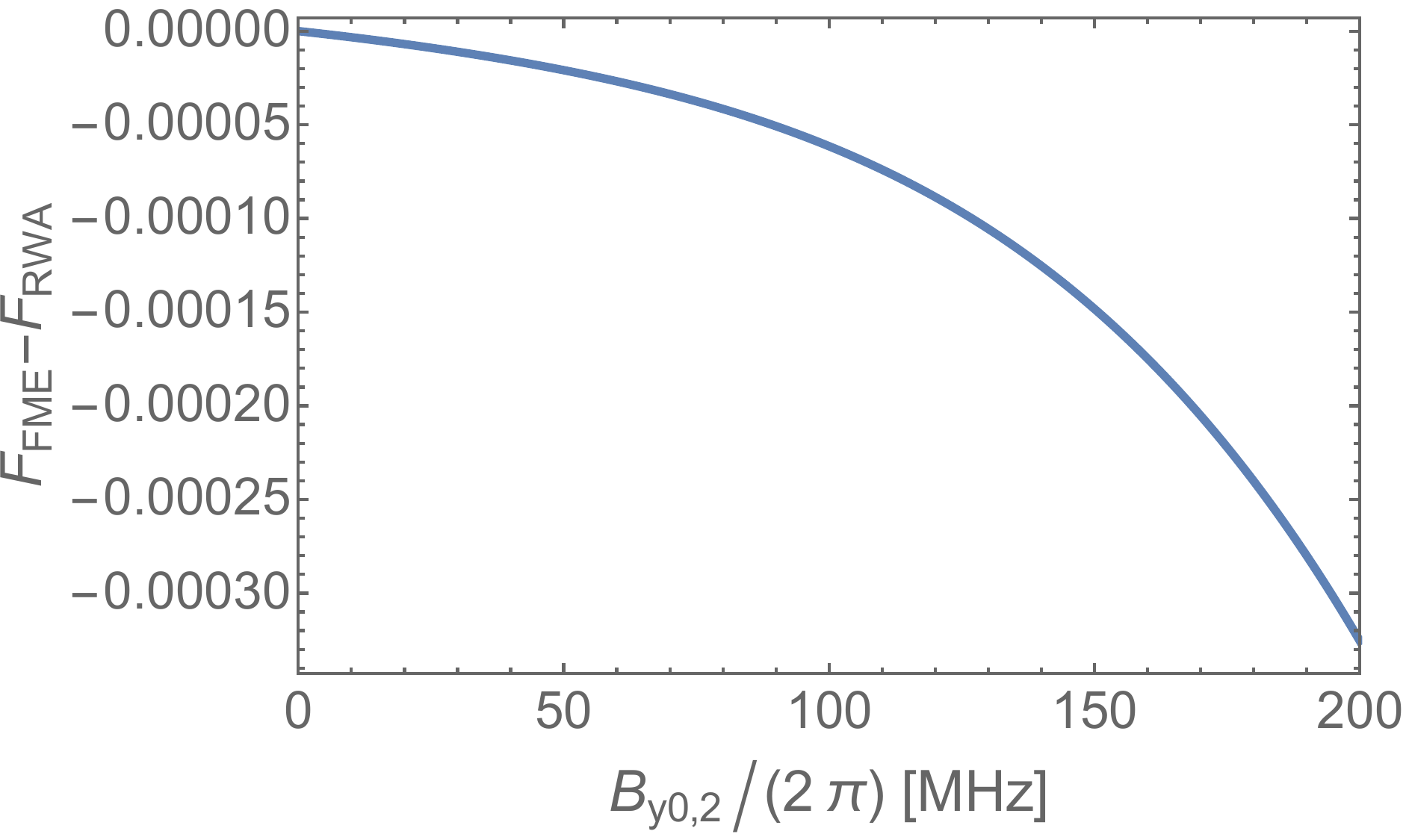}
	\caption{Impact of the first Floquet-Magnus correction on the neighbor qubit fidelity for increasing $B_{y0,2}$ when performing a Y gate.}
	\label{FME}
\end{figure}

To investigate the validity of the RWA we plot the impact of the first order Floquet-Magnus correction~\cite{Bukov_2015,Blanes_2009,Moore_1990,Mostafazadeh_1997} on the fidelity depending on $B_{y0,2}$ of the neighbor qubit~2 when performing a Y gate on qubit~1 in Fig.~\ref{FME} with $B_{z,1} = (2\pi)\, 18.483$~GHz, $\Delta B_z = (2\pi)\, 0.2$~GHz, $B_{y1,1} = (2\pi)\, 200$~MHz, and $B_{y1,2} = 0.4 \, B_{y1,1}$.
Indeed RWA shows to be a valid approximation in our case and we can determine the fidelity impairment for a given value of $B_{y0,2}$.

\section{Charge noise analysis}\label{appNoise}
For a charge noise analysis assuming fluctuations in the EDSR driving fields $B_{y1,1}$ and $B_{y1,2}$ matrix elements of the neighbor qubit Hamiltonian causing Rabi oscillations are considered. First order corrected diagonal and off-diagonal elements are given by
\begin{widetext}
    \begin{align}
    &U_{\text{diag}} = \cos\left(\frac{\tilde{\Omega}_{\text{NN}}}{2}t\right)
    \mp \frac{i \delta\omega_z}{\tilde{\Omega}_{\text{NN}}}\sin\left(\frac{\tilde{\Omega}_{\text{NN}}}{2}t\right) + \left| \frac{\Omega_{\text{NN}}}{2 \tilde{\Omega}_{\text{NN}}} \left( -\frac{t}{2} \mp \frac{i \delta\omega_z}{\tilde{\Omega}_{\text{NN}}^2} \right) \sin\left( \frac{\tilde{\Omega}_{\text{NN}}}{2}t \right) 
    \mp\frac{ i\delta\omega_z \Omega_{\text{NN}} t}{4 \tilde{\Omega}_{\text{NN}}^2}\cos\left( \frac{\tilde{\Omega}_{\text{NN}}}{2}t \right) 
    \right|\,
    \delta B,\\
    &U_{\text{off-diag}} = \pm \frac{\Omega_{\text{NN}}}{\tilde{\Omega}_{\text{NN}}} \sin\left(\frac{\tilde{\Omega}_{\text{NN}}}{2}t\right)
    + \left| \pm \frac{\delta\omega_z^2}{2 \tilde{\Omega}_{\text{NN}}^3} \sin\left( \frac{\tilde{\Omega}_{\text{NN}}}{2}t \right) \mp \frac{\Omega_{\text{NN}}^2 t}{4 \tilde{\Omega}_{\text{NN}}^2}\cos\left( \frac{\tilde{\Omega}_{\text{NN}}}{2}t \right) \right|\,
    \delta B.
\end{align}
\end{widetext}

\begin{figure}[t]
    \centering
    \includegraphics[width=0.49\textwidth]{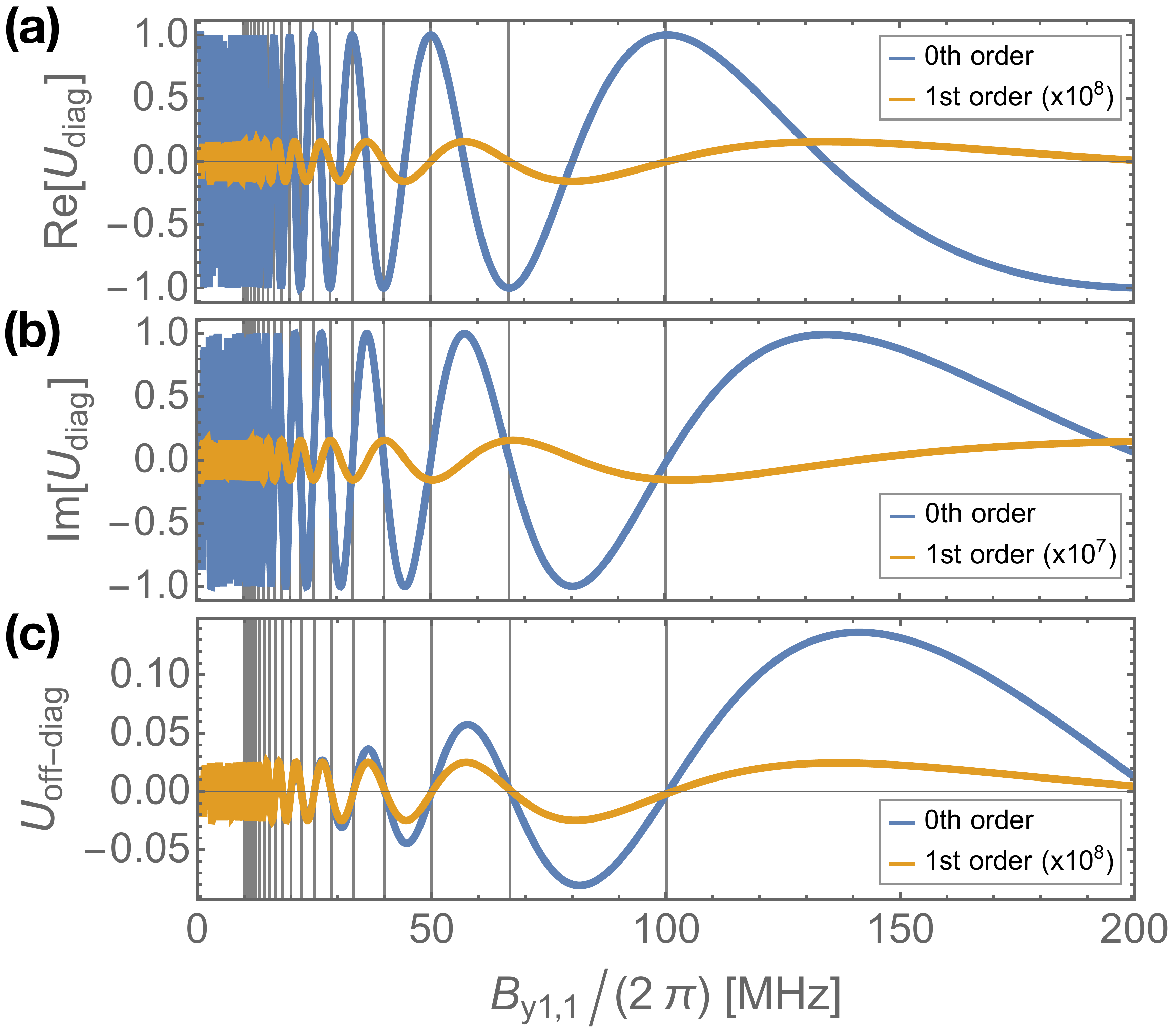}
    \caption{First order noise correction to \textbf{(a)} real and \textbf{(b)} imaginary diagonal and \textbf{(c)} off-diagonal Rabi terms of the neighboring qubit for the Y gate with $\delta \omega_z = \Delta B_z = (2\pi)\, 0.2$ GHz and $\alpha = 0.4$, where vertical lines mark $B_{y1,1}$ values matching the synchronization condition.}
    \label{noiseAppY}
\end{figure}

First order corrections of the diagonal and off-diagonal terms are shown in Fig. \ref{noiseAppY} for the Y gate, where for clarity reasons noise amplitudes are multiplied with factors $10^8$ and $10^7$, respectively. Ideally the absolute value of the real diagonal part should be 1 while the imaginary diagonal and the off-diagonal part should be zero, which is fulfilled for the zeroth order terms at synchronization. Apparently, the first order real diagonal part is cancelled and the off-diagonal part is almost zero when $B_{y1,1}$ fulfills the crosstalk synchronisation condition. The remaining imaginary diagonal noise contribution relative to zeroth order is of approximately one magnitude smaller than the off-resonant contribution, which is why the overall noise contribution is nearly minimal at synchronization as shown in the infidelity plot in Section~\ref{section:noise}, where for first order zero mean Gaussian noise with $\sigma_{B}$ of 0~MHz, $(2\pi)\,$10~MHz and $(2\pi)\,$15~MHz the total infidelity is plotted. Similar behaviour is obtained for the CNOT gate where the first order noise amplitude relative to the zeroth order amplitude of the imaginary part is even smaller compared to the off-diagonal term.
For the values used and proposed in this paper, crosstalk synchronization is hardly sensitive to first order noise via EDSR field fluctuations, and thus shows to be robust.

\bibliography{bibliography}

\end{document}